\magnification=1200
\baselineskip=20 pt

\def\eijk{\epsilon_{ijk}}

\def\nubl{\bar{\nu}_L}
\def\ebl{\bar{e}_L}
\def\dbri{\bar{d}_{Ri}}
\def\ucrj{u^c_{Rj}}
\def\c{\cos\theta}
\def\s{\sin\theta}

\def\gpr{g^{\prime}_R}
\def\gppl{g^{\prime\prime}_R}
\def\ubcli{\bar{u}^c_{Li}}
\def\dbcli{\bar{d}^c_{Li}}
\def\e{\epsilon}

\centerline{\bf Higgs induced light leptoquark-diquark
mixing  and proton decay}

\vskip 1 true in
\centerline{\bf Uma Mahanta}
\centerline{\bf Mehta Research Institute}
\centerline{\bf Chhatnag Road, Jhusi}
\centerline{\bf Allahabad-211019, India}
\centerline{\it email:mahanta@mri.ernet.in}
\vskip .4 true in

\centerline{\bf Abstract}

In low energy phenomenology to avoid the strong constraints of proton
decay it is usually assumed that light ($\approx $ 250 Gev) leptoquarks 
couple only to quark-lepton pairs and light diquarks couple only to 
quark pairs. In this paper we present two specific examples where 
the higgs induced mixing between leptoquarks and diquarks through trilinear 
interaction terms reintroduces the troublesome
couplings and gives rise to proton decay. 
The bound on the unknown parameters of
this scenario that arise from proton life time has also been derived.

\vfill\eject

Leptoquarks (LQ) [1] and diquarks (DQ) [2, 3]
 are colored scalar or vector particles 
that carry baryon numbers of $\pm {1\over 3}$ and $\pm {2\over 3}$ 
respectively. They occur naturally in many extensions of the standard
model (SM) e.g. superstring inspired grand unified models based on
$E(6)$ [3], technicolor models [2] and composite models [1]
 of quarks and leptons.
If LQs and DQs couple both to quark-lepton pairs and quark pairs
then they lead to too rapid proton decay. To be consistent with the 
proton life time such LQs and DQs must have a mass of the order of
$(10^{12}-10^{15})$ Gev. To avoid this strong constraint and to make them
relevant for low energy phenomenology below 1 Tev it is usually assumed
that LQs couple to quark-lepton pairs but not to quark pairs
and DQs couple to quark pairs but not to quark-lepton pairs. However in the
presence of the SM higgs doublet $\phi $ the low energy effective Lagrangian
will also contain a trilinear Higgs-LQ-DQ interaction term. After EW
symmetry breaking (EWSB) this interaction term will induce a mixing between 
the  LQ and the DQ which reintroduces the troublesome Yukawa couplings
for LQ and DQ. If the physical LQ and DQ do not have the same mass then
they lead to proton decay. There can be several different kinds [1-3]
of LQ's and DQ's. However in this report we shall not go into a 
detailed discussion of every LQ-DQ mixing that can possibly occur.
Rather we shall give two give specific examples
of Higgs-LQ-DQ interaction term and the mixing between the LQ and DQ
that takes place after EWSB. For one particular case we have also calculated
the mixing angle in terms of the dimensional coupling that measures the 
strength of the Higgs-LQ-DQ interaction term. Finally we have estimated the
proton decay rate in terms of the unknown parameters of our scenario
and have derived the constraint that the parameters of the Lagrangian
must satisfy in order to be consistent with proton lifetime.

Consider the SM to be extended by a light ($\approx $ 250 Gev ) chiral 
leptoquark
D and a light chiral diquark S with the following assignments under 
$SU(3)_c\times SU(2)_l\times U(1)_y$: $D\sim (3^*, 2, -{1\over 6})$
and $S\sim (3^*, 1, {1\over 3})$. The low energy effective Lagrangian of 
this extended scenario will contain besides the SM Lagranigian all possible
renormalizable and gauge invariant interaction terms between D, S and the SM
fields. Of particular importance for this work are the Yukawa like couplings
of D and S to the SM fermions given by the following Lagrangian

$$\eqalignno{L_y&=g_R{\bar l}_L d_R D+g^{\prime }\epsilon_{ijk} 
\bar {d}_{Ri}u^c_{Rj}S_k
 +h.c.\cr
&=g_R (\bar {\nu}_L D_1+\bar {e}_L D_2)d_R+g^{\prime}_R
\eijk \dbri\ucrj S_k+h.c. &(1)\cr}$$

Here $D_1$ and $D_2$ are the isospin up and down components of D.
i, j, k are the color indices. Note first that the leptoquark
 D couples only to quark-lepton pair and the diquark S
 couples only to quark pair which is required so that they do not 
lead to proton decay. 
Second both interaction terms 
are invariant under the SM gauge group $SU(3)_c\times SU(2)_l\times U(1)_y$.
Finally the couplings of D and S are both chiral in nature i.e. they
couple to quark fields of a particular chirality only. The Yukawa Lagrangian
$L_y$ also implies that D and S should carry baryon numbers of $-{1\over 3}$
and ${2\over 3}$ respectively so that $L_y$ conserves baryon number.
This kind of structure of the Yukawa couplings could arise depending
on the gauge symmetry of the high energy theory and its chiral matter
representation.
Besides the Yukawa couplings the low energy Lagrangian for describing 
physics much below the compositeness scale or grand unified scale $\Lambda$
must also contain all possible renormalizable and gauge invariant
interaction terms between $\phi $, D and S. In principle such interaction
terms cannot be neglected. They could arise from the scalar potential
 of the high energy theory after the gauge symmetry of the high energy
theory breaks down
into $SU(3)_c\times SU(2)_l\times U(1)_y$.
 The gauge invariant trilinear 
interaction term between $\phi $, D and S is given by the Lagrangian
$$L_s=k_1(D^+\phi_c)S+h.c.=k_1{v+h\over \sqrt{2}}D^+_1S +h.c.\eqno(2)$$
Here $\phi_c$ is the charge conjugated Higgs doublet that gives mass to the up 
quarks in SM. The unknown mixing
 parameter $k_1$ carries the dimension of mass.
After EW symmetry breaking this interaction term will lead to mixing 
between $D_1$ and S. Note that $D_1$ and S carries the same charge
and color assignments which is necessary so that they could mix after EWSB.
The above higgs-LQ-DQ interaction term violates both baryon number and
lepton number by one unit ($\delta B=-\delta L =1$). Here we are 
implicitly assuming that the breaking of baryon number symmetry
is first communicated to the scalar sector of the high energy
theory which in turn communicates it to other sectors e.g the
Yukawa sector. 
 Recall that in the
SM baryon number conservation is realized as an accidental global
symmetry [4]. By that we mean that given the particle content and the 
gauge group $SU(3)_c\times SU(2)_l\times U(1)_y$ of the SM
it is not possible to write a renormalizable and gauge invariant 
interaction term that violates baryon number. No adhoc global symmetry is
required to explain the near abscence of proton decay.
However as we have seen above that if we keep the gauge group the same but
allow additional particles like color triplet scalars then it is possible
to write down a renormalizable and gauge invariant interaction term that
violates baryon number. 

We shall now show that the mixing between $D_1$ and S induced by $\phi$
can lead to proton decay in the ``$\nu$ + any channel'' where any
refers to a positively charged non-strange meson. Consider the full
 scalar potential involving D and S
$$\eqalignno{V(D, S)&= \mu_1^2D^+D+\mu_2^2S^+S+\lambda_1(D^+D)^2+
\lambda_2(S^+S)^2\cr
&+\lambda_1^{\prime}(D^+D)(\phi^+\phi )+\lambda_2^{\prime}(S^+S)(\phi^+\phi )
+(k_1D^+\phi_cS+h.c.)&(3)\cr}$$

The quartic scalar interactions are not relevant to our present work.
$\mu_1^2$ and $\mu_2^2$ are the mass parameters associated with the 
gauge eigenstates D and S. After EWSB the trilinear interaction term 
between $\phi$, D and S induces a mixing between $D_1$ and S.
It can be shown that the eigenvalues of the mass squared matrix 
$M^2_{D_1S}$ are given by
$$\lambda_+=M^2_{D_1}={1\over 2}({\mu^2_1+\mu^2_2})+{1\over 2}
\sqrt{ \e^2+2k_1^2v^2}.\eqno(4a)$$
and
$$\lambda_-=M^2_S={1\over 2}({\mu^2_1+\mu^2_2})-{1\over 2}
\sqrt{ \e^2+2k_1^2v^2}.\eqno(4b)$$
 where $\e =\mu^2_1-\mu^2_2$.
The physical states correponding to $\lambda_{\pm}$ are given by
$D_1^{\prime}=D_1\cos\theta -S\sin\theta $ and $S^{\prime}=D_1\sin
\theta +S\cos\theta $. Here $\cos\theta ={\sqrt {2}k_1 v\over
[2k_1^2v^2+(\sqrt{\e^2+2k_1^2v^2}-\e )^2]^{1\over 2}}$.

 The Yukawa couplings of the LQ and the DQ
written in terms of mass eigenstates are given by
$$\eqalignno{L_y&= g_R[\nubl (D_1^{\prime}\c +S^{\prime}\s +\ebl D_2]d_R\cr
&+\eijk \gpr\dbri\ucrj [S_k^{\prime}\c -D^{\prime}_{1k}\s +h.c.]
 &(5)\cr}$$.
The above Yukawa Lagrangian does lead to proton decay in the ``$\nu $
+ any'' channel via the exchange of D and S particles. The effective four
fermion Lagrangian for this decay is given by 
$$L_{eff}=g_R \gpr \s \c \eijk (\nubl d_{Rk}) (\bar {u}^c_{Rj} d_{Ri})
({1\over M_S^2}-{1\over M_D^2})+h.c.\eqno(6) $$

This is the effective Lagrangian at a scale $M^2=M^2_D\approx M^2_S \approx
(250 Gev)^2$. In order to use it for proton decay it has to be renormalized
down to a scale $\mu^2=(1 Gev )^2$ under the unbroken QCD and 
EM interactions. The EM corrections are small because the coupling itself
is small. The QCD corrections are not that large either because
$\ln{M^2\over \mu^2}$ is not large in our case. It can be shown that the 
proton decay rate arising from the above effective Lagrangian is given by
$$\Gamma (p\rightarrow \nu_e +any )\approx {1\over 17 \pi^2 R^3}G_{eff}^2
m_q^2 A^2 ({M\over \mu})\eqno(7)$$.

Here $A({M\over \mu})$ includes the effects of renormalization group
evolution from M to $\mu$. We shall assume it to be of order one.
 $R={3\over 4}$ fm, $m_q\approx {1\over 3}$ Gev and
$G_{eff}={1\over 2\sqrt {2}}g_R \gpr \s\c ({1\over M^2_S}-{1\over M^2_D})$.
The present lower bound on $\tau (p\rightarrow \nu+any )$ is 25$\times 
10^{30}$ yr [5]. It is clear from the expression of $G_{eff}$ that the tight
constraint of proton lifetime can be satisfied either with
$\sin\theta \ll 1$ or $\cos\theta \ll 1$. It can be shown that for
$\s\ll 1$ we require $k_1v\ll \sqrt {2}\e$. On the other hand for
$\c \ll 1$ we need $\sqrt {2}k_1v\ll (\sqrt {\e^2+2k_1^2v^2}-\e )$. 
For $M_S= 200$ Gev and $M_D=300$ Gev [6]
 the product combination
$g_R\gpr \s\c $ must be less than $3\times 10^{-26}$ in order that the
decay rate in eqn(7) is consistent with the proton lifetime.
 If we assume that the Yukawa couplings
$g_R$ and $\gpr$ are of the order of .1, then for  $\s\ll 1$
 we must adjust the parameters of the Lagrangian so as to satisfy
$k_1v<4\times 10^{-24}\e $. On the other hand for $\c\ll 1$
the parameters must be adjusted so as to satisfy 
$\sqrt{2}k_1v<3\times 10^{-24} (\sqrt {\e^2+2k_1^2v^2}-\e )$.
These constraints may not be stable against radiative corrections
requiring extreme fine tuning to each order in loop expansion.
We would like to reiterate that there can be several different species of
LQ's as well as DQ's.
 However the mixing between $D_1$ and S and hence
proton decay occurs only for specific
$SU(3)_c\times SU(2)_l\times U(1)_y$ assignments of the LQ and the DQ.
Unless these assignments are achieved there is no mixing between the LQ
and DQ and hence no contribution to proton decay.

Having shown that higgs induced mixing between the doublet leptoquark
 D and the 
singlet diquark S does lead to proton decay we shall now present another 
concrete example where the same phenomenon also takes place.
An EW triplet diquark $T_{ak}$ (a is the $SU(2)_l$ index and k the color index)
 can also mix with the doublet leptoquark D and 
contribute to both proton decay and neutron decay. Let the 
$SU(3)_c\times SU(2)_l\times U(1)_y$ assignments of $T_{ak}$ be given 
by (3, 3, $-{1\over 3}$). 
The triplet diquark $T_{ak}$ can couple to LH quark pair according
to the following Lagrangian:
$$\eqalignno{L^{\prime}_y&=\gppl \eijk \bar{q}^c_{Li}\tau_a i\tau_2
T_{ak}+h.c.\cr
&=\gppl \eijk [-\sqrt{2}\ubcli u_{Lj}T_{-k} +\sqrt{2}\dbcli d_{Lj}T_{+k}
+(\ubcli d_{Lj}+\dbcli u_{Lj})T_{0k}]&(6)\cr}$$
where $T_{+k}={T_{1k}+iT_{2k}\over \sqrt{2}}$, $T_{-k}={T_{1k}-iT_{2k}
\over \sqrt{2}}$ and $T_{0k}=T_{3k}$. $T_{+k}$ and $T_{-k}$ carry
electromagnetic charges of ${2\over 3}$ and $-{4\over 3}$ units
respectively. Note that $T_{-k}$ is not the antiparticle of $T_{+k}$
since they carry different electromagnetic charges.
The gauge invariant Higgs-LQ-DQ interaction term in this case will be 
given by
$$\eqalignno{L^{\prime}_s&=k_2 (D^+\tau_a\phi_c )T^*_a+h.c.\cr
&=k_2{v+h\over \sqrt{2}}[D_2^+T^*_+ +D^+_1 T_0^* +h.c.]&(7)\cr}$$
The mixing between $D_1$ and $T_0^*$leads to $p\rightarrow \pi^+\nu$
and the mixing between $D_2$ and $T^*_+$ leads to $n\rightarrow \pi^+e^-$
both of which violate baryon number.

The low energy effective Lagrangian will also contain trilinear
Higgs-LQ-LQ and Higgs-DQ-DQ interaction terms. Afetr EWSB these terms 
give rise to mixing between different multiplets of LQ and DQ.
Such interaction terms do not violate baryon number but they could violate
lepton number. Of particular importance is the Higgs-LQ-LQ ineraction
for a pair of chiral leptoquarks belonging to different weak SU(2)
multiplets. The mixing between  two different 
 LQ multiplets will give rise to
helicity unsuppressed contribution to $\pi^-\rightarrow e^-\bar{\nu}_e$
The helicity suppressed SM contribution to
$\pi^-\rightarrow e^-\bar{\nu}_e$ is in excellent agreement with the
 experimental data. Therefore any helicity unsuppressed contribution 
arising from new physics must be strongly constrained. This leads to
stringent bounds on the mixing parameter between the two LQ multiplets.
 The mixing between
different LQ multiplets can also lead to majorana mass matrix
for neutrinos.
A complete discussion of these topics can be found in Ref. [7] and 
will not be taken up here.

To conclude in this report we have shown that the usual assumption of low 
energy phenomenology that LQ's couple only to q-l pairs and DQ's couple 
only to quark pairs is not sufficient to stabilize the proton.
We have given two specific examples where the 
Higgs-LQ-DQ interaction term   induces a mixing between 
the LQ and DQ after EWSB. This mixing
reintroduces the troublesome couplings for the LQ and the DQ and leads to 
proton decay. We find that in order to be consistent with the bound on
proton lifetime the parameters of the Lagrangian must either satisfy
$k_1v<4\times 10^{-24}\e $ or $\sqrt{2}k_1v<3\times 10^{-24}(\sqrt{\e^2
+2k_1^2v^2}-\e )$. In deriving these constraints we have assumed that
$g_R\approx \gpr \approx .1 $, $M_{D_1}=300 $ Gev and $M_S$=200 Gev.
These constraints may not be stable against radiative corrections
requiring extreme fine tuning of the parameters to each order
in loop expansion.
This naturalness problem can perhaps be resloved by means of some 
new symmetry that remains unbroken in the low energy theory and prevents
the Higgs-LQ-DQ interaction term from occuring in the effective
Lagrangian.

\centerline{\bf References}

\item{1.} W. Buchmuller Acta Phys. Austriaca, Suppl 27, 517 (1985); W. 
Buchmuller and D. Wyler, Phys. Lett. B 177, 377 (1986).

\item{2.} E. Farhi and L. Susskind, Phys. Rep. 74, 277 (1981); K. Lane and
M. V. Ramanna, Phys. Rev. D 44, 2678 (1991).

\item{3.} J. Hewett and T. Rizzo, Phys. Rep. 183, 193 (1989).

\item{4.} A. Cohen, A. Nelson and D. Kaplan, Phys. Lett. B 388, 588 (1996).

\item{5.} Review of Particle Properties Euro. Phy. Jour. C3, 50 (1998).

\item{6.} The values of $M_S$ and $M_D$ assumed by us are consistent with 
the latest bounds on LQ and DQ masses: Euro. Phys. Jour. C3, 260 (1998);
B. Abbot et al. (DO Collaboration), Phys. Rev. Lett. 80, 2051 (1998);
F. Abe et al. (CDF Collaboration), Phys. Rev. D 55, R5263 (1997).

\item{7.} M. Hirsch et al., Phys. Rev. D 54, 4207(1996); Phys. Lett. 
B 378, 17 (1996); U. Mahanta, MRI-PHY/990930, hep-ph/9909518.

\end